# Nature of Science in the classroom: empirical insights into investigating physics

Daniel Trugillo Martins Fontes
University of Sao Paulo
ORCID: 0000-0002-4741-2067

**Abstract**

In this article we present and discuss a didactic activity that combines epistemic aspects of the nature of science (NOS) and a simple, relatively low-cost experiment on electromagnetic induction. The aim of the activity is for students to gain a deeper understanding of the relationship between theory and experiment. The results of the application showed that the students not only improved their communication skills within the groups, but also exercised creativity when facing a real and unknown problem. In addition, interactions with the students revealed that they assimilated some of the consensual aspects of the nature of science, namely: there is no universal scientific method, scientists use imagination and belief in scientific practice, and theory is not a consequence of experimental observation and vice versa. In relation to pedagogical aspects, students have improved their communicative abilities within groups, exercised creativity in problem-solving, and assumed responsibility in the development of the activity. Although the activity was carried out with high school students, it also has potential to be applied in higher education. Finally, various possibilities for expansion and modifications to the activity are discussed, aiming to further enrich the learning experience.

**1. Introduction**

In recent years, research in science education – particularly in physics education – has increasingly focused on how scientific concepts are taught in light of contemporary concerns [1]. Simultaneously, we have witnessed growing instances of unfounded criticism toward science and scientific practice, along with the spread of misinformation, fake news, and pseudoscience [2]. In response to this troubling trend, national and international science education journals have dedicated special issues to exploring ways to reflect on and address the dissemination of misinformation and scientific skepticism, as well as their impact on science teaching. One of the most pressing issues raised is the lack of classroom attention to the nature of science itself. A lack of deep reflection and insufficient emphasis on the characteristics of scientific inquiry are contributing factors in the spread of pseudoscience and science denialism [3–5].

Although this broader scenario stems from multiple factors that extend beyond formal education, the physics education research community can contribute by developing classroom activities that highlight key aspects of the nature of science [6]. As a result, the nature of science (NOS) has become an increasingly relevant topic in academic discourse and the subject of substantial literature. As Moura [7] succinctly puts it, NOS refers to the set of elements involved in the construction, validation, and organization of scientific knowledge. This includes both epistemic dimensions – such as scientific methods and the interplay between theory and experimentation – and non-epistemic ones, like the influence of sociocultural, religious, and political factors on the acceptance of scientific ideas. Broadly speaking, fostering an understanding of NOS is considered one of the core objectives of science education [8].

Within this context, the purpose of this study is to present a teaching activity designed to engage students in discussing and experiencing epistemic aspects of NOS, particularly by fostering a more critical understanding of the relationship between theory and experiment. Teachers can adopt a variety of strategies to introduce NOS in the classroom. In this work, we chose to investigate a simple and inexpensive experiment involving free fall and electromagnetism. It is worth highlighting that *Física na Escola* – Brazil's most established journal dedicated to physics education, especially in basic education – frequently publishes experimental activities [9], reflecting the ongoing interest and relevance of this approach within the educational community it serves.

Overall, the main goal of this study is to offer support for physics teachers aiming to address epistemic aspects of NOS through experimental practice. The central focus is to encourage reflection on how experiments can be used to explain theoretical concepts – particularly those that are typically accepted only when presented in algebraic form. One example is the concept of the electromagnetic field, which is described in the literature and textbooks in various ways, such as "electromagnetic substance," "a kind of aura," "a field of invisible lines," or "a disturbance in space" [10, p. 1084].

## 2. Brief Overview of the Physics

The experiment discussed in this study involves two key physical phenomena: the free fall of a body and electromagnetic effects. When a magnet is dropped into a conducting tube, an effect occurs that can be described by Lenz's Law, which states that the direction of any magnetic induction effect is such that it opposes the cause that produced it [11]. This law can be derived from Faraday's Law of Electromagnetic Induction and is closely tied to the principle of energy conservation. A qualitative representation of electromagnetic induction inside the coil is shown in Figure 1.

**Figure 1**: Illustration of a magnetic object moving through a coil. Image not to scale[2]

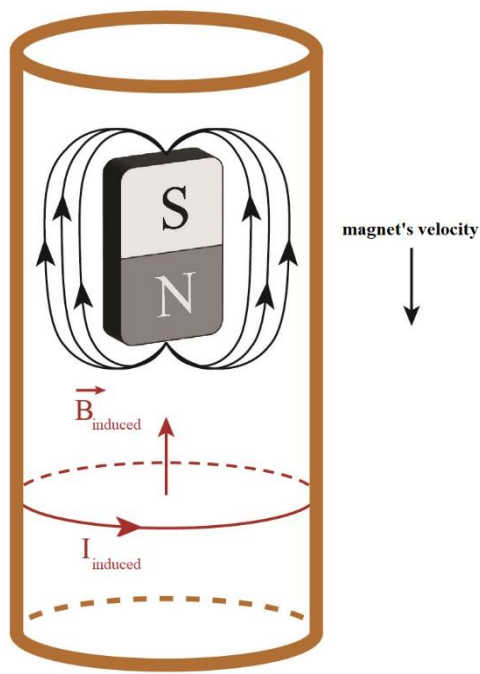

As shown in Figure 1, a magnet moves inside a conducting tube. This motion causes a change in magnetic flux, which in turn induces an electric current that opposes the change. The induced current generates a magnetic field oriented in the opposite direction of the flux variation caused by the moving magnet. Careful measurements of this phenomenon show that the magnet reaches terminal velocity shortly after being released [12]. This occurs because the faster the magnet moves, the greater the change in magnetic flux – leading to a stronger induced field that resists the motion of the magnet, thus slowing it down.

That said, the details of electromagnetic induction – including the magnitude and direction of the induced current and magnetic field – require significant abstraction, theoretical knowledge of electromagnetism, and a level of mathematical proficiency beyond what is expected of most high school students, including those involved in this activity. As such, we acknowledge that electromagnetic induction is not a trivial phenomenon. However, in order to provoke critical thinking about the relationship between theory and experiment, as well as the role of observation and inference in scientific understanding, we deliberately chose a physical phenomenon whose effects are not visually evident. Experiments that, for instance, involve turning a light on or off, or observing whether an object or gauge moves, would not be as effective for the type of inquiry we aimed to promote.

---

[2] Due to an editorial oversight, Figure 1 was not included in the published version.

## 3. The Didactic Activity
### 3.1. Specific Objectives

There is extensive literature on learning objectives in science education, both in terms of inquiry –based learning and the Nature of Science (NOS). Previous studies have proposed guiding questions such as: "What are the limits and possibilities of scientific knowledge?" [13, p. 8]. And how can a basic education teacher "engage students' inherent curiosity and desire to interact by using science's own tools – observation, hypothesis formulation, experimentation, analysis, and so on?" [14, p. 34]. Although these questions were posed some time ago, they remain highly relevant and are aligned with the current general competencies of basic education, as defined by Brazil's National Common Curricular Base (BNCC). As such, they informed the formulation of the following specific objectives:

- To stimulate reflection, hypothesis generation, discussion, and the construction of explanations about what can and cannot be directly observed;
- To promote argumentation grounded in empirical evidence;
- To provide hands-on experience with scientific procedures typical of knowledge production, including observation, questioning, systematization, and peer discussion;
- To encourage critical reflection on the relationship between theory and experimentation.

With these objectives in place, the activity was implemented in a regular public high school located on the coast of São Paulo state. The class included 25 third-year students (ages 17 to 18), evenly distributed by gender, and took place during the evening in a 90-minute session. Although the school is situated in the city's central area, it functions more like a neighborhood school, primarily serving students from nearby communities. The school does have a science laboratory, but it is rarely used by teachers. In 2021, the school's Basic Education Development Index (IDEB) was 4.5 – very close to the average for the state of São Paulo.

### 3.2. Materials and Procedures

The experiment involved 11 different materials, as shown in Figure 2. The teacher used a notebook to record observations, while the students used their personal cell phones as stopwatches.

**Figure 2**: Material available to students

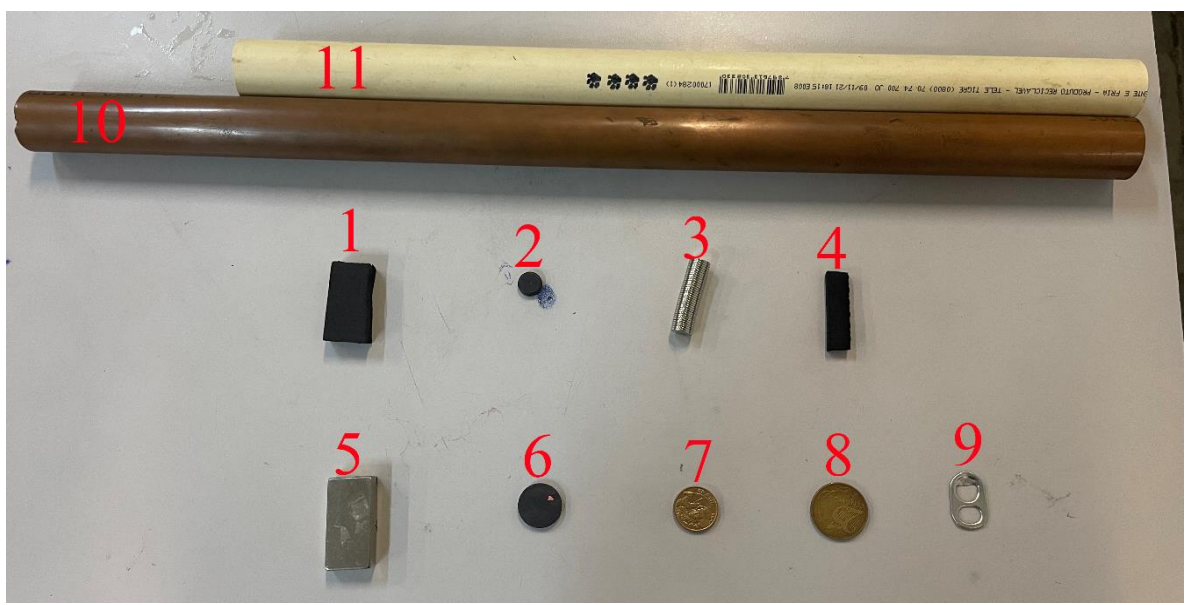

The items in Figure 2 include: pieces of rubber (items 1 and 4), neodymium magnets (3 and 5), ferrite magnets (2 and 6), coins (7 and 8), an aluminum seal (9), a 51-centimeter-long copper cylinder (10), and a PVC plastic cylinder (11). The rubber pieces and neodymium magnets were intentionally made to have very similar dimensions (see pairs 1–5 and 3–4 in Figure 2). The experimental procedure allowed considerable freedom: students could choose which materials to test and in what order, leaving those decisions to each group.

Students were instructed to hold the tubes vertically and as steadily as possible. While one student dropped the object into the tube, another measured the fall time. It's important to note that neodymium magnets are relatively fragile and can chip or break if they hit the floor. For this reason, students were advised to place a hand or piece of clothing beneath the tube when working with these magnets.

### 3.3. In the classroom

Before beginning the hands-on investigation, the teacher led a whole-class discussion to activate students' prior knowledge about magnetic materials. This initial moment was key to gathering information that helped guide the teacher's questioning strategy. It also helped establish a participatory and motivating environment for student engagement [15]. The overall structure of the lesson was inspired by Santiago's recent work [16], which employed an empirical-inductivist approach to exploring magnetic fields. Table 1 outlines the steps followed during the activity.

**Table 1**: Summary of lesson stages and didactic flow.

| | |
|---|---|
| Stage 1 (15 min) | **Exploring students' prior knowledge of magnetism and epistemic aspects of NOS.** The teacher and students engaged in dialogue and collaboratively created a mind map on the board centered on the concept of "theory." Each student who contributed a term came to the front to write it on the board and explain its relevance. To probe prior conceptions of magnetism, the teacher used guiding questions adapted from Souza Filho et al. [15, p. 33], such as: What is a magnet? What are its characteristics? What kinds of materials are attracted to magnets?. |
| Stage 2 (5 min) | **Problem.** A key NOS question was introduced to guide reflection: *How does a scientist know whether an experiment confirms a theory?* This question was written on the board to serve as a reference point for the lesson. |
| Stage 3 (10 min) | **Presentation of materials.** A brief overview of the materials was provided, without revealing which objects were magnetic. The experimental procedure was explained—holding the tube vertically, recording the object and time, etc. As students interacted with the materials and questions emerged, the teacher offered further clarification. Students were also advised to keep their phones away from the neodymium magnets to avoid potential damage. |
| Stage 4 (40 min) | **Group exploration and knowledge building.** Students were free to manipulate the materials and formulate hypotheses based on their observations. The teacher monitored group progress and discussions, stepping in only when needed. Students were encouraged to reflect on their methods—for example, discussing how switching who held the stopwatch might influence their timing. |
| Stage 5 (20 min) | **Knowledge synthesis.** In this final stage, the teacher collected the materials and students' notes, in which they had described their investigative process, stated hypotheses, and indicated whether those hypotheses were confirmed or refuted. A class discussion helped synthesize the findings and return to the initial problem. The session concluded with the recreation of the earlier mind map, now enriched by the students' experimental experience. |

**Figure 3**: Mind maps illustrating connections between theory and scientific experimentation. The central term "theory" was the only one written by the teacher.

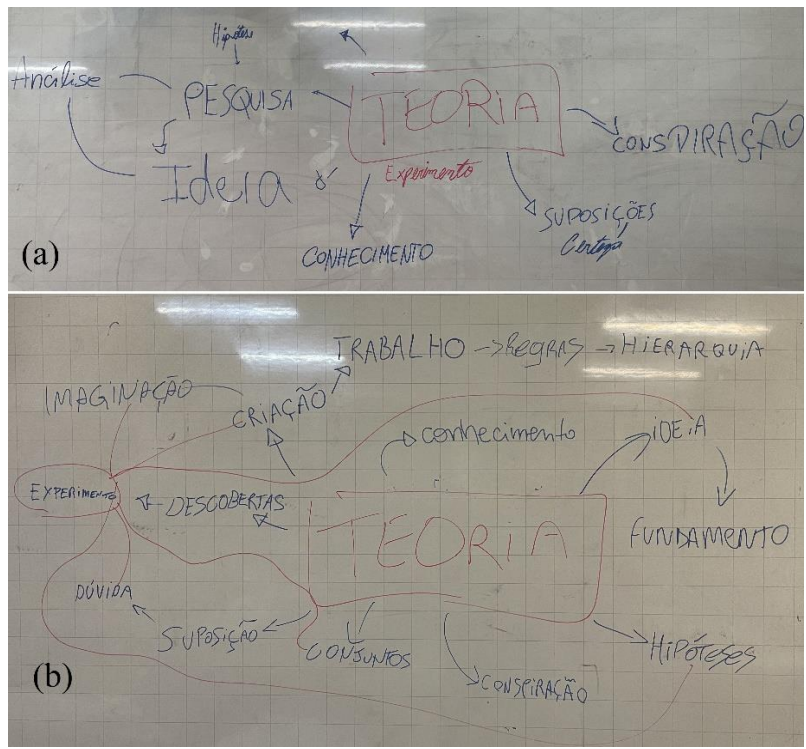

Figure 3 presents some of the mind maps created by students during Stage 1. It was surprising to see how rich the students' mind maps were in terms of conceptual content and associations with their understanding of scientific theory. Several terms appeared across both maps, such as knowledge, hypotheses, conspiracy, and idea. However, in one class, the concept of experiment appeared on its own, without clear connections to other terms, while in the other, it was linked to ideas like imagination, creation, hypothesis, idea, and doubt. These concepts were later revisited during the discussion in Stage 5.

Following Santiago's approach [16], this activity adopted an empirical-inductivist methodology, relying on repeated observation and investigation of a phenomenon, partially mirroring the procedures of scientific research. In this specific context, electromagnetic phenomena cannot be directly observed, which created an opportunity for students to construct knowledge through their own actions while developing core investigative skills. Figure 4 captures key moments from Stages 4 and 5.

**Figure 4**: Scenes from Stages 4 and 5. (a) Students investigating the problem; (b) Students discussing the meaning of their results; (c) Example of hypotheses – highlighted in red – produced by one of the groups.

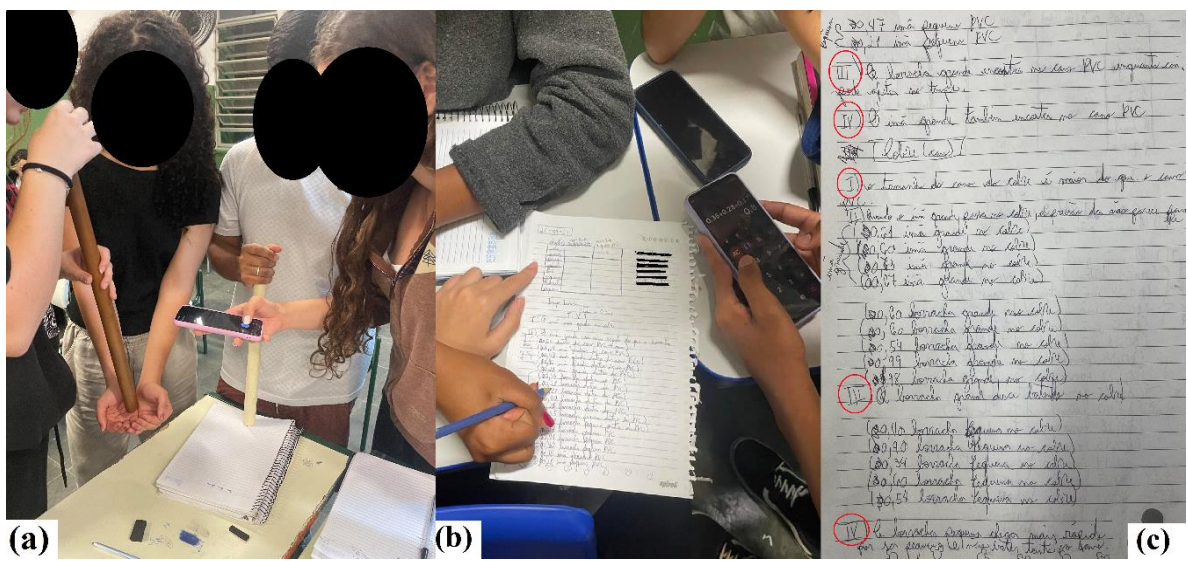

In Figure 4c, it's clear that the students' hypotheses emerged progressively throughout the data collection process. The first two hypotheses generated by the group shown were: "the magnet doesn't stick to the copper" and "the magnet falls faster than the rubber." Other highlighted hypotheses included: "the large rubber hits the PVC pipe as it falls, which might affect the timing"; "the large magnet also touches the PVC pipe"; "when the large magnet passes through the copper, the hand feels some pressure"; and "the small rubber falls faster because it's smaller and doesn't hit the pipe as much." These student statements – regardless of their physical accuracy – suggest that the activity successfully encouraged thoughtful connections between the experimental materials and the behavior of a magnetic object falling through a conducting tube. They also illustrate varying levels of abstraction in the students' observations, both direct (*e.g.*, the magnet is not attracted to copper) and indirect (*e.g.*, sensations felt in the hand holding the coil as the magnet passes through).

## 4. General Discussion

Through the guiding questions and hands-on activity described, our aim was to encourage and challenge students to engage in scientific inquiry that would help them develop and apply experimental skills. Throughout this process, students were invited to refine their understanding of the natural world, actively participating in the construction of their own knowledge. The teacher plays a key role in this context, acting as a mediator between scientific knowledge and classroom knowledge, fostering an environment conducive to learning, and offering support and guidance as needed.

The proposed experiment proved highly effective – not only did students engage in the task itself, but they also took part in defining the research question, formulating hypotheses, discussing ideas with the teacher, and exploring different ways of collecting and interpreting data. Early in the activity, students' comments revealed that their intuitive understanding of electromagnetism was tied to more familiar phenomena, such as light or the attraction between two magnets. While this made the task of teaching electromagnetic induction more challenging, it also opened valuable space for reflection, discussion, and scientific argumentation, allowing students to confront the relationship between theory and experimental observation. In this exploration of the "invisible", students were no longer passive recipients of information – they became active participants in the meaning-making process. One illustrative moment occurred when, after being asked whether an object could fall faster inside one of the tubes, a student replied: "No, because if it did, it would have more energy out of nowhere, which is weird…then everyone would have free energy, right?"

Reflections like this emphasize the teacher's central role in facilitating the activity, guiding students toward critical thinking and encouraging their active engagement in the learning process. As Paulo Freire put it [17, p. 35, my translation]:

> As [the teacher] dialogues with students, he must draw their attention to points that may be unclear or overly simplistic, always problematizing them. Why? What do you mean? What connection do you see between what you just said and what your classmate 'A' said? Is there a contradiction? Why?

Problematization means identifying the questions students need to grapple with. It aligns with a dialogical approach that provokes curiosity and reflection, stimulating the development of critical thinking in the construction of scientific knowledge. Under the teacher's guidance, students are exposed to situations that require them to go beyond naïve inductivism – the assumption that observation is inherently objective and always leads to unquestionable conclusions [7] – and toward a more critical understanding of how science works.

### 4.1. Student Engagement and Interaction

By the end of Stage 4, it became evident that students strongly believed that "something" was influencing the large magnet as it fell through the coil, causing it to descend more slowly. They questioned whether this was "really true," adopting a common student stance that places the teacher as the ultimate holder of truth. To encourage reflection on the Nature of Science (NOS), I chose to step into the role of a scientist and explained the concept of electromagnetic induction, referencing Faraday's and Lenz's laws – while intentionally avoiding appeals to authority. One notable observation was that, without any prompting, the students repeated their experiments several times to calculate an average fall time or arrive at more reliable results. Table 2 shows some of the recorded averages, as documented by one of the groups.

Table 2: Average fall times recorded by one group.

| Material | Avg. fall time (coil) | Avg. fall time (PVC cylinder) |
|---|---|---|
| Large magnet (item 5, Fig. 2.) | 0.63s | 0.33s |
| Large rubber (item 1, Fig. 2.) | 0.47s | 0.34s |
| Small magnet (item 3, Fig. 2.) | 0.29s | 0.31s |
| Small rubber (item 4, Fig. 2.) | 0.29s | 0.33s |

It is known that the theoretical fall time $t$ of an object in free fall near Earth's surface can be calculated using the simplified formula:

$$t = \sqrt{\frac{2h}{g}}$$

where $h$ is the height (in this case, 0.51m – the length of the coil) and $g = 9{,}8$ m/s². This gives an approximate theoretical fall time of 0.32s. As seen in Table 2, the average fall time for the large rubber piece was significantly higher than for the small one. This happened because the large rubber's diameter was very close to the inner diameter of the coil, causing occasional friction against the tube walls if students didn't keep it properly aligned. This did not pose a serious issue during the lesson, as students already understood that an object's mass does not affect its fall time.

In Stage 5, we introduced a new variable into the analysis: human reaction time. I shared data indicating that average reaction time among adolescents ranges from 0.24 to 0.97 seconds [18]. Based on this, I posed a new question: How can you be sure that the large magnet is falling more slowly due to a physical phenomenon and not because of your own reaction time when stopping the stopwatch? Even after considering this, some students were adamant that they had "definitely seen" the large magnet fall more slowly. This led us to another line of questioning: If you were living in Aristotle's time, would your observations have been the same? Since students had already been introduced to Aristotelian concepts of motion, the example resonated well[3]. It allowed us to explore the role of theory in shaping observation – emphasizing that scientists don't experiment in a social vacuum; rather, they are guided by expectations, imagination, and assumptions, as the students themselves pointed out (see Fig. 3b). Observation is always theory-laden, and scientific theories are shaped and reshaped over time within broader socio-historical contexts.

---

[3] Broadly speaking, in Aristotelian physics, heavy objects (e.g., stones) fall toward the center of the Earth through a natural motion, as they are made of the same substance as the planet itself. In contrast, Newtonian physics explains this motion as the result of a gravitational force acting between bodies with mass. Thus, while the experience of observing falling bodies is the same, the theories that support them are different.

During our final discussion, students began asking how they could be more confident in their conclusions. Suggestions included using "more objects to test," "a machine-triggered timer," or "a tool to measure electric current." These insights suggest that students had begun to grasp that scientific knowledge evolves with the availability of more precise tools and data – especially when comparing objects that are identical in every way except their magnetic properties. This shift in thinking – such as exploring induction not just by fall time but also via current measurement – demonstrates engagement with key dimensions of NOS that the activity set out to explore.

Ultimately, we concluded that the activity helped students meaningfully engage with and internalize core epistemic aspects of the nature of science. The experience also validated the effectiveness of the methodology, as it supported students in deepening their understanding of scientific practice – even when they already had some prior knowledge. Additionally, the collaborative nature of the activity encouraged students to work together, strategize, and take ownership of the collective process of scientific discovery and knowledge negotiation.

### 4.2. Suggestions for Adaptations and Future Activities

Schools, teachers, and students operate in diverse contexts and hold different expectations when it comes to teaching and learning. Nevertheless, incorporating critical reflection on how scientific knowledge is produced is a shared responsibility that should be actively encouraged within the school community. The activity presented in this study was inspired by gaps and suggestions identified in the literature on physics and science education, many of which were referenced earlier. In that spirit, we offer a set of possible adaptations aimed at addressing both epistemic and non-epistemic dimensions of the Nature of Science (NOS). These suggestions are intended to serve as practical resources for educators and researchers and to support the development of future instructional sequences.

1. This activity was also implemented with first-year high school students. In that context, the aim was not to explore electromagnetism, but rather to focus on scientific practices and the concept of free fall. Interestingly, the session lasted significantly longer with this group, as students were eager to conduct multiple tests using all of the items shown in Figure 2. This did not occur with senior students, who quickly realized that items like the coins and the aluminum seal were distractors that didn't contribute meaningfully to the investigation. For this reason, we suggest that if the activity is used with younger students, the materials be limited to the copper coil and the rubber and neodymium magnets (items 1,

3, 4, and 5 in Fig. 2). There is no need to include ferrite magnets or the PVC tube in this case, since the focus shifts from exploring electromagnetism to investigating free fall and scientific inquiry through collaboration, error, and experimentation. Additionally, teachers may choose to reduce the degree of freedom in the investigation. For example, a procedural guide could be provided, specifying the order of experiments so that only one variable is changed at a time. This strategy allows students to make clearer comparisons. It's also worth noting that using longer tubes helps minimize measurement errors and makes the magnetic drag effect more apparent.

2. Measuring short time intervals with a handheld stopwatch can lead to imprecise data and may cause students to doubt the results. The influence of human reaction time introduces significant uncertainty. To address this, teachers might incorporate digital tools that allow for more accurate time measurements. For instance, the free and easy-to-use Audacity app can be used to measure fall times based on sound. When the object exits the tube and hits the surface, a spike appears on the audio waveform. A brief introductory lesson on time measurement can help students interpret their results, especially given that fall times are recorded to the hundredth of a second.[4]

3. Teachers may also choose to implement this activity after students have formally studied electromagnetism. This approach may be more appropriate for advanced high school classes or undergraduate courses. While the core of the activity remains unchanged, some stages may be carried out differently. For instance, during Stage 5, students could be asked to explain the electromagnetic phenomenon inside the tube, draw schematic diagrams, and describe the application of Faraday's and Lenz's laws in their own words. They could also be encouraged to classify materials (e.g., magnetic, conductive, insulating) and relate these categories to the outcomes observed during the activity.

4. It may also be worthwhile to introduce the class to the article by Silveira and Ostermann [19], which questions the notion that scientific laws can be straightforwardly derived from experimental results. This reading provides valuable insights and may help educators address philosophical questions that arise during the activity. For example, "If we had more experimental data, would we be able to explain what's happening with certainty?" As a complementary resource, the article by Assis and Neves [20] presents a historical debate in modern cosmology and offers a compelling case for exploring how scientific knowledge evolves in relation to empirical evidence.

5. Teachers could also incorporate dynamics that foster critical thinking about the non-epistemic aspects of NOS. In the original activity, all materials were freely distributed to student groups. An alternative approach would be to ask students to submit a written justification for why they wish to test a particular item before gaining access to it. This mirrors actual research practices, where access to scientific instruments or datasets is often contingent on clearly articulated hypotheses or research objectives. A review panel represented by the teacher or even classmates could approve or deny the request based on the strength of the reasoning. There are many theoretical frameworks that could support this variation, but Thomas Kuhn [21] stands out for his accessible language and introductory appeal.

---

[4] I would like to thank the anonymous reviewer for suggesting this contribution.

6. Stage 5 could be modified to further challenge the idea of science as a collection of fixed truths. For instance, students could submit anonymous written hypotheses about what happens when objects fall through the tubes. Each group would draft an argument and submit it in a sealed envelope, which would then be shuffled and redistributed. Groups would evaluate the anonymous arguments and assign them scores. The highest-rated argument would be deemed the "scientific truth" by peer consensus. In a follow-up session, the teacher, playing the role of a scientist with new evidence, could introduce alternative interpretations. The class, again acting as a scientific community, would vote on whether to update the accepted explanation. This activity allows students to reflect on both the personal and collective dimensions of science, and to understand how scientific consensus is constructed, evaluated, and revised over time. It also highlights the social processes that underlie scientific validation, key elements of the non-epistemic side of NOS.

These adaptations and extensions can enrich students' understanding of science, making it more reflective, participatory, and aligned with the complexities of real scientific practice. While any implementation must take into account the students' level, learning goals, and available resources, the overarching aim should remain the same: to encourage discussion and critical reflection about how science operates and how scientific knowledge is constructed.

**5. Conclusion**

This study set out to present and reflect on a teaching activity designed to challenge commonly held assumptions about the scientific process and the nature of science. The approach enabled students to engage in inquiry while contributing their own perspectives, resulting in an experience more closely connected to the real world. We believe this process fosters a deeper, more meaningful understanding of scientific practice, allowing students to take an active role in the construction of knowledge.

Developing an understanding of the nature of science is essential for preparing students to critically engage with scientific issues throughout their lives, regardless of their chosen fields. We acknowledge that there are many strategies available to teachers to support this goal, and the activity we propose is just one possible path. What matters most is cultivating an environment in which students feel empowered to question, explore, and critically examine how science works. This activity also encouraged students to reflect critically on fundamental questions, such as: (i) whether a single, universal scientific method exists, (ii) whether theory is simply the result of experimentation and vice versa, (iii) and whether scientific knowledge relies solely on direct observation, or also on inference and interpretation grounded in complex and often elusive evidence.

Throughout the activity, students not only improved their communication skills within their groups but also exercised creativity in problem-solving and took ownership of completing the tasks. Finally, we echo and expand on the reflection by Alcântara and Moura [22] concerning the role of the history of science in physics education, advocating for a broader integration of philosophy of science with hands-on experimentation in the classroom.